\documentclass[aps,prd,draft,showpacs]{revtex4}

\begin{document}

\title{On the properties of the vacuum expectation value in $R_{\xi }$ gauge
and $\overline{R_{\xi }}$ gauge}
\author{Chungku Kim}
\affiliation{Department of Physics, College of Natural Science, Keimyung
University, Daegu 705-701, KOREA}
\date{\today}

\begin{abstract}
We have investigated the gauge dependence of the vacuum expectation
value(VEV) both in the $R_{\xi }$ and the $\overline{R_{\xi }}$
gauge in the $\overline{MS}$ scheme. We have found that, in case of the $R_{\xi }$ gauge, the gauge
dependence of the VEV should be modified due to the presence of the parameter
in the gauge function that should be identified as a VEV in the broken
symmetry phase. However the pole mass remains gauge independent.
\end{abstract}
\maketitle



\affiliation{Department of Physics, College of Natural Science, Keimyung
University, Daegu 705-701, KOREA }

In case of the gauge theories where the gauge symmetry is broken
spontaneously, the Lagrangian density in the broken symmetry phase $L_{BS}$
is given from the one in the symmetric phase $(L_{SYM})$\ as 
\begin{equation}
L_{BS}(H,v)=\left[ L_{SYM}(\Phi )\right] _{\Phi =H+v},
\end{equation}
where $v$ is the vacuum expectation value(VEV). Then mixing term between the
gauge boson and the Goldstone boson emerges in the quadratic part of the
Lagrangian in the broken symmetry phase which can complicate the
perturbation calculations for the physical processes and in order to remove
this mixing term, the $R_{\xi }$ gauge\cite{R-ksi} is used usually. However, in the
broken symmetry phase, the following properties were taken for granted.

(i)The gamma function of the VEV is same as that of the Higgs field$(\gamma
_{v}=\gamma _{\phi })$.

(ii) $\xi \frac{\partial v}{\partial \xi }=C_{\xi }(v)$ where $C_{\xi }(v)$
is the function given by Nielsen\cite{Nielsen}.

(iii)The effective potential $V(\phi )$ at $\phi =v$ and the pole mass is
gauge independent.

Recently, the renormalization properties of the VEV in case of the $R_{\xi }$
gauge were investigated\cite{VEV-1}\cite{VEV-2} in the minimal
renormalization(MS) scheme. It was shown that the gamma function of the VEV($%
\gamma _{v}$) have an additional contributions originating from the additive
renormalization of the scalar field and as a result, was found to be
different from that of the scalar field$(\gamma _{\phi })$ in $R_{\xi }$
gauge and the property(i) does not hold in this gauge. In this paper, we
investigate the property (ii) and (iii) in case of broken phase of both the $%
R_{\xi }$ gauge and the $\overline{R_{\xi }}$ gauge\cite{Kas} which
also remove the mixing terms in the quadratic part of the Lagrangian.
Recently it was shown that the VEV obtained from the one-loop effective
potential $\overline{R_{\xi }}$ gauge satisfies the property(i)\cite
{Kim1}. In order to see the difference between the $R_{\xi }$ and $\overline
{R_{\xi }}$ gauges, let us consider the scalar electrodynamics with
the Lagrangian density in the symmetric phase

\begin{eqnarray}
L_{SYM}(\Phi _{1},\Phi _{2},A_{\mu }) &=&\frac{1}{4}F_{\mu \nu }F_{\mu \nu }+%
\frac{1}{2}(\partial _{\mu }\Phi _{1}+gA_{\mu }\Phi _{2})^{2}+\frac{1}{2}%
(\partial _{\mu }\Phi _{2}-gA_{\mu }\Phi _{1})^{2}+\frac{1}{2}m^{2}(\Phi
_{1}^{2}+\Phi _{2}^{2})+\frac{\lambda }{24}(\Phi _{1}^{2}+\Phi _{2}^{2})^{2}
\nonumber \\
&&+\frac{1}{2\xi }f(\Phi _{1},\Phi _{2},A_{\mu })^{2}+\overline{c}\frac{%
\delta f(A_{\mu }^{\theta },\Phi _{2}^{\theta })}{\delta \theta }c+counter%
\text{ }terms,
\end{eqnarray}
where 
\[
F_{\mu \nu }=\partial _{\mu }A_{\nu }-\partial _{\upsilon }A_{\mu }.
\]
$L_{SYM}(\Phi _{1},\Phi _{2},A_{\mu })$ has the local $O(2)$ symmetry is
given by 
\begin{equation}
\delta _{\theta }A_{\mu }=\partial _{\mu }\theta ,\text{ }\delta _{\theta
}\Phi _{1}=\theta \Phi _{2},\text{ }\delta _{\theta }\Phi _{2}=-\theta \Phi
_{1},
\end{equation}
except the gauge-fixing function $f(A_{\mu },\Phi )$ and the resulting ghost
term $\overline{c}\frac{\delta f(A_{\mu }^{\theta },\Phi _{2}^{\theta })}{%
\delta \theta }c.$ When $m^{2}$ becomes negative$,\Phi _{1}$ develops a VEV
and the Lagrangian density in the broken phase can be obtained by
substituting $\Phi _{1}$ with $H+v$ where $H$ is the Higgs field having a
vanishing VEV. As a result, mixing terms $-gvA_{\mu }\partial _{\mu }\Phi
_{2}$ emerges which can complicate the perturbation calculations for the
physical processes. In the case of the $\overline{R_{\xi }}$ gauge, $%
f(\Phi _{1},\Phi _{2},A_{\mu })$ is given by 
\begin{equation}
f(\Phi _{1},\Phi _{2},A_{\mu })=\partial _{\mu }A_{\mu }-\xi g\Phi _{1}\Phi
_{2},
\end{equation}
in the symmetric phase and when we replace $\Phi _{1}$ by $H+v$ in the
broken symmetric phase, mixing terms $-gv\Phi _{2}\partial _{\mu }A_{\mu }$
emerges which can remove the above mixing term. In case of the $R_{\xi }$
gauge, $f(\Phi _{1},\Phi _{2},A_{\mu })$ is given by 
\begin{equation}
f(\Phi _{1},\Phi _{2},A_{\mu })=\partial _{\mu }A_{\mu }-\xi gu\Phi _{2},
\end{equation}
in the symmetric phase and in the broken symmetry phase, we should choose $%
u=v$ besides replacing $\Phi _{1}$ by $H+v$ in order to remove the mixing
term $-gvA_{\mu }\partial _{\mu }\Phi _{2}$. This means that in case of the $%
R_{\xi }$ gauge, the parameter $u$ in the symmetric phase of the Lagrangian
should be identified as VEV in the broken symmetry phase and hence we
will consider these two cases separately.

\subsection{\protect\bigskip Gauge fixing with no parameter in the symmetric phase
related to VEV ( $\overline{R_{\xi }}$ gauge)}

Let us first consider the case of the $\overline{R_{\xi }}$ gauge
where no parameter of the Lagrangian in the symmetric phase is related to
the VEV. Then the VEV is obtained from the renormalized effective potential
of the symmetric phase ($V_{SYM}$) as 
\begin{equation}
\left[ \frac{\partial V_{SYM}(\phi )}{\partial \phi }\right] _{\phi =v}=0.
\end{equation}
The renormalized effective potential of the symmetric phase ($V_{SYM}$)
satisfy the renormalization group(RG) equation as\cite{RG} 
\begin{equation}
DV_{SYM}+\gamma _{\phi }\phi \frac{\partial V_{SYM}}{\partial \phi }=0,
\end{equation}
where 
\begin{equation}
D=\mu \frac{\partial }{\partial \mu }+\beta _{g}\frac{\partial }{\partial g}%
+\beta _{\lambda }\frac{\partial }{\partial \lambda }+\beta _{m^{2}}\frac{%
\partial }{\partial m^{2}}+\beta _{\xi }\frac{\partial }{\partial \xi }.
\end{equation}
Then by applying the operator $D$ to Eq.(6) and by using Eq.(7), we obtain 
\begin{eqnarray}
0 &=&D\left[ \frac{\partial V_{SYM}}{\partial \phi }\right] _{\phi
=v}=\left[ D\frac{\partial V_{SYM}(\phi )}{\partial \phi }\right] _{\phi
=v}+(Dv)\left[ \frac{\partial ^{2}V_{SYM}(\phi )}{\partial \phi ^{2}}\right]
_{\phi =v}  \nonumber \\
&=&-\gamma _{\phi }\left[ \frac{\partial V_{SYM}(\phi )}{\partial \phi }%
+\phi \frac{\partial ^{2}V_{SYM}(\phi )}{\partial \phi ^{2}}\right] _{\phi
=v}+(Dv)\left[ \frac{\partial ^{2}V_{SYM}(\phi )}{\partial \phi ^{2}}\right]
_{\phi =v}=(Dv-\gamma _{\phi }v)\left[ \frac{\partial ^{2}V_{SYM}(\phi )}{%
\partial \phi ^{2}}\right] _{\phi =v},
\end{eqnarray}
and as a result we obtain\cite{Kim2} 
\begin{equation}
Dv=\gamma _{\phi }v.
\end{equation}
This shows that when the Lagrangian in the symmetric phase $(L_{SYM})$ is
related to the VEV, it follows that $\gamma _{v}=\gamma _{\phi }.$ Now,
consider the Nielsen identity\cite{Nielsen} in case of the symmetric phase
of the gauge theory given by 
\begin{equation}
\xi \frac{\partial V_{SYM}(\phi )}{\partial \xi }+C_{\xi }(\phi )\frac{%
\partial V_{SYM}(\phi )}{\partial \phi }=0.
\end{equation}
Then, by applying $\xi \frac{\partial }{\partial \xi }$ to Eq.(6) and by
using Eq.(11), we obtain 
\begin{eqnarray}
0 &=&\xi \frac{\partial }{\partial \xi }\left[ \frac{\partial V_{SYM}(\phi )%
}{\partial \phi }\right] _{\phi =v}=-\left[ \frac{\partial C_{\xi }(\phi )}{%
\partial \phi }\text{ }\frac{\partial V_{SYM}(\phi )}{\partial \phi }+C_{\xi
}(\phi )\frac{\partial ^{2}V_{SYM}(\phi )}{\partial \phi ^{2}}\right] _{\phi
=v}+\xi \frac{\partial v}{\partial \xi }\left[ \frac{\partial
^{2}V_{SYM}(\phi )}{\partial \phi ^{2}}\right] _{\phi =v}  \nonumber \\
&=&(\xi \frac{\partial v}{\partial \xi }-C_{\xi }(v))\left[ \frac{\partial
^{2}V_{SYM}(\phi )}{\partial \phi ^{2}}\right] _{\phi =v},
\end{eqnarray}
and as a result we obtain\cite{Fraser} 
\begin{equation}
\xi \frac{\partial v}{\partial \xi }=C_{\xi }(v).
\end{equation}
By using Eq.(11) and (13), we can see the well known that the effective
potential is gauge independent at $\phi =v$ as 
\begin{equation}
\xi \frac{d}{d\xi }V(v)=\xi \frac{\partial v}{\partial \xi }\frac{\partial
V(v)}{\partial v}-C_{\xi }(v)\frac{\partial V(v)}{\partial v}=0.
\end{equation}
Now, by applying $\xi \frac{\partial }{\partial \xi }$ to the equation that
defines the pole mass $M^{2}$ as 
\begin{equation}
\left[ \frac{\partial ^{2}\Gamma _{SYM}}{\partial \phi ^{2}}\right] _{\phi
=v,\text{ }p^{2}=-M^{2}}=0,
\end{equation}
where $\Gamma _{SYM}$ is the effective action in the symmetric phase and by
using Eq.(11), we obtain 
\begin{eqnarray}
0 &=&\xi \frac{\partial }{\partial \xi }\left[ \frac{\partial ^{2}\Gamma
_{SYM}}{\partial \phi ^{2}}\right] _{\phi =v,\text{ }p^{2}=-M^{2}}=\xi \frac{%
\partial v}{\partial \xi }\left[ \frac{\partial ^{3}\Gamma _{SYM}}{\partial
\phi ^{3}}\right] _{\phi =v,\text{ }p^{2}=-M^{2}}+\xi \frac{\partial M^{2}}{%
\partial \xi }\left[ \frac{\partial ^{3}\Gamma _{SYM}}{\partial \phi
^{2}\partial (p^{2})}\right] _{\phi =v,\text{ }p^{2}=-M^{2}}  \nonumber \\
&&-\left[ \frac{\partial ^{2}C_{\xi }(\phi )}{\partial \phi ^{2}}\text{ }%
\frac{\partial \Gamma _{SYM}}{\partial \phi }+2\text{ }\frac{\partial C_{\xi
}(\phi )}{\partial \phi }\text{ }\frac{\partial ^{2}\Gamma _{SYM}}{\partial
\phi ^{2}}+C_{\xi }(\phi )\text{ }\frac{\partial ^{3}\Gamma _{SYM}}{\partial
\phi ^{3}}\right] _{\phi =v,\text{ }p^{2}=-M^{2}}.
\end{eqnarray}
By using Eqs.(6),(13) and (15), we can see that only the second term of the
first line survives in Eq.(16) and as a result, we obtain the gauge
independence of the pole mass\cite{Nielsen}\cite{Fraser}
\begin{equation}
\xi \frac{\partial M^{2}}{\partial \xi }=0,
\end{equation}
which has been verified explicitly up to two-loop order in case of the $%
\overline{R_{\xi }}$ gauge in Ref.\cite{Kas}.

\subsection{Gauge fixing with a parameter in the symmetric phase related
to VEV($R_{\xi }$ gauge)}

In order to see the changes in the $R_{\xi }$ gauge, let us first note that
since $\xi ,u$ are gauge parameters in the symmetric phase the corresponding
Nielsen identities are given as\cite{Zeit} 
\begin{equation}
x\frac{\partial V_{SYM}}{\partial x}+C_{x}(\phi )\frac{\partial V_{SYM}}{%
\partial \phi }=0\ (x=\xi ,\text{ }u).
\end{equation}
As explained below Eq.(5),$\ $we should take the gauge parameter \ $u$ as
VEV in the broken symmetry phase in order to remove the cross terms in the
quadratic Lagrangian and as a result the effective potential in the broken
symmetry phase $V_{BS}$ is given by 
\begin{equation}
V_{BS}(h,v)=\left[ V_{SYM}(\phi ,u)\right] _{u=v,\phi =h+v}.
\end{equation}
Then the minimum condition for the VEV becomes 
\begin{equation}
0=\left[ \frac{\partial V_{BS}(h,v)}{\partial v}\right] _{h=0}=\left[ \frac{%
\partial V_{SYM}}{\partial \phi }+\frac{\partial V_{SYM}}{\partial u}\right]
_{\phi =u=v}=(1-\frac{C_{u}(v)}{u})\left[ \frac{\partial V_{SYM}}{\partial
\phi }\right] _{\phi =u=v}=0,
\end{equation}
which gives the same condition as given in Eq.(6). In applying the RG
operator $D$ to the minimum condition for the VEV, note that the RG function
for $\phi $ in MS scheme in the $R_{\xi }$ gauge is different from the usual
one and have an inhomogeneous term coming from the tadpole diagrams
generated from the $\xi gu\Phi _{2}\partial _{\mu }A_{\mu }$ term of the $%
R_{\xi }$ gauge\cite{Willey}. In order to see this, consider the one-loop
effective potential in the symmetric phase is given by\cite{Kim1}\cite{Fraser}\cite{Willey}
\begin{equation}
V_{1}=-\frac{\hbar }{2}Tr\ln D_{H}^{-1}-\frac{\hbar }{2}Tr\ln D_{G}^{-1}-%
\frac{\hbar }{2}Tr\ln X^{-1}+\hbar Tr\ln D_{g}^{-1},
\end{equation}
where 
\begin{equation}
D_{H}^{-1}=p^{2}+m_{H}^{2},
\end{equation}
\begin{equation}
D_{G}^{-1}=p^{2}+m_{G}^{2}+\xi g^{2}u^{2},
\end{equation}
\begin{equation}
X_{\mu \nu }^{-1}=(p^{2}+m_{A}^{2})(\delta _{\mu \nu }-\frac{p_{\mu }p_{\nu }%
}{p^{2}})+\frac{D(p^{2})}{m_{G}^{2}+\xi m_{A}^{2}}\frac{p_{\mu }p_{\nu }}{%
\xi p^{2}},
\end{equation}
and 
\begin{equation}
D_{g}^{-1}=p^{2}+m_{g}^{2},
\end{equation}
\begin{equation}
D(p^{2})=p^{4}+(m_{G}^{2}\text{ }+2\xi gum_{A})p^{2}+\xi
m_{A}^{2}(m_{G}^{2}+\xi g^{2}u^{2})\equiv (p^{2}+m_{+}^{2})(p^{2}+m_{-}^{2}),
\end{equation}
with 
\begin{equation}
m_{H}^{2}=m^{2}+\frac{\lambda }{2}\phi ^{2},m_{G}^{2}=m^{2}+\frac{\lambda }{6%
}\phi ^{2},\text{ }m_{A}=g\phi ,\text{ }m_{g}^{2}=\xi gum_{A}.
\end{equation}
By performing the one-loop momentum integral in $D\equiv 4-2\varepsilon $
dimension\cite{integral}, we obtain the $\frac{1}{\varepsilon }$ divergence
as 
\begin{equation}
V_{1,div}=\frac{\hbar }{16\pi ^{2}\varepsilon }\{\frac{1}{4}m_{H}^{4}+\frac{1%
}{4}m_{+}^{4}+\frac{1}{4}m_{-}^{4}+\frac{3}{4}m_{A}^{2}-\frac{1}{2}%
m_{g}^{4}\}.
\end{equation}
The $m_{+}^{4}+m_{-}^{4}$ term in Eq.(28) contains the $\phi $ and $\phi
^{3} $ term which arises due to the tadpole diagrams mentioned above. Since
these terms were absent in the tree Lagrangian and hence can not be removed
solely from the multiplicative renormalization, and we should choose 
\begin{equation}
\phi _{B}=\sqrt{Z_{\phi }}(\phi +u\text{ }Z_{u}),
\end{equation}
which gives the RG function of the form 
\begin{equation}
\mu \frac{\partial \phi }{\partial \mu }=\gamma _{\phi }\phi +\delta \text{ }%
u.
\end{equation}
From Eq.(28), one can check that $Z_{u}=$ $\frac{1}{16\pi ^{2}\varepsilon }%
\xi g^{2}$ and by noting that $\mu \frac{\partial g}{\partial \mu }%
=-\varepsilon g+...$ , we have $\delta =\frac{1}{8\pi ^{2}}\xi g^{2}$ at one
loop order. Then the general form of the RG equation in $R_{\xi }$ gauge can
be written as 
\begin{eqnarray}
&&DV_{SYM}+u\text{ }\gamma _{u}\frac{\partial V_{SYM}}{\partial u}+(\gamma
_{\phi }\phi +\delta \text{ }u)\frac{\partial V_{SYM}}{\partial \phi } 
\nonumber \\
&=&DV_{SYM}+(\gamma _{\phi }\phi +\delta \text{ }u-C_{u}\gamma _{u})\frac{%
\partial V_{SYM}}{\partial \phi }=0,
\end{eqnarray}
where $D$ is given in Eq.(8) and we have used Eq.(18) to obtain the second
line of above equation.

When we apply the operator $D$ to the minimum condition as in Eq.(6), we
should note the fact that VEV comes out not only from $\phi $ but also from $%
u$ so that 
\begin{eqnarray}
\ \text{\ }0 &=&D\left[ \frac{\partial V_{SYM}}{\partial \phi }\right]
_{\phi =u=v}=\left[ D\frac{\partial V_{SYM}}{\partial \phi }\right] _{\phi
=u=v}+(Dv)\left[ \frac{\partial ^{2}V_{SYM}}{\partial \phi ^{2}}+\frac{%
\partial ^{2}V_{SYM}}{\partial u\partial \phi }\right] _{\phi =u=v} 
\nonumber \\
&=&-\left[ \gamma _{\phi }\frac{\partial V_{SYM}}{\partial \phi }+(\gamma
_{\phi }\phi +\delta \text{ }u-C_{u}\gamma _{u})\frac{\partial ^{2}V_{SYM}}{%
\partial \phi ^{2}}\right] _{\phi =u=v}+(Dv)\left[ \frac{\partial ^{2}V_{SYM}%
}{\partial \phi ^{2}}+\frac{\partial ^{2}V_{SYM}}{\partial u\partial \phi }%
\right] _{\phi =u=v}.
\end{eqnarray}
Then, by using Eq.(20) and $x=u$ case of Eq.(18), we obtain 
\begin{equation}
\left[ \frac{\partial ^{2}V_{SYM}}{\partial u\partial \phi }\right] _{\phi
=u=v}=-\frac{C_{u}(v)}{v}\left[ \frac{\partial ^{2}V_{SYM}}{\partial \phi
^{2}}\right] _{\phi =u=v},
\end{equation}
and from Eq.(32) we obtain 
\begin{equation}
Dv=\frac{(\gamma _{\phi }+\delta )v-C_{u}(v)\gamma _{u}}{1-C_{u}(v)/v}.
\end{equation}
As $\gamma _{\phi }$, $\delta ,C_{u}$ and $\gamma _{u}$ are one-loop
quantities, the resulting one-loop terms in the above equation are $(\gamma
_{\phi }+\delta )v$ and by noting that $\gamma _{\phi }=\frac{1}{16\pi ^{2}}%
(3-\xi )g^{2}$\cite{RG-1}\cite{RG-2} and with the result $\delta =\frac{1}{%
8\pi ^{2}}\xi g^{2}$ given above, we can see that this agrees with\cite
{VEV-1}\cite{VEV-2}. Beyond the two loop order, our result of $Dv$ will be
different from \cite{VEV-1}\cite{VEV-2} due to the $C_{u}(v)$ dependent
terms. Now, let us consider the gauge dependence of the VEV in $R_{\xi }$
gauge. By applying $\xi \frac{\partial }{\partial \xi }$ to Eq.(6) which is
same condition as (20) and by using Eq.(18) and Eq.(33), we obtain 
\begin{eqnarray}
\xi \frac{\partial }{\partial \xi }\left[ \frac{\partial V_{SYM}(\phi )}{%
\partial \phi }\right] _{\phi =u=v} &=&-\left[ \frac{\partial C_{\xi }(\phi )%
}{\partial \phi }\text{ }\frac{\partial V_{SYM}(\phi )}{\partial \phi }%
+C_{\xi }(\phi )\frac{\partial ^{2}V_{SYM}(\phi )}{\partial \phi ^{2}}%
\right] _{\phi =u=v}+\xi \frac{\partial v}{\partial \xi }\left[ \frac{%
\partial ^{2}V_{SYM}(\phi )}{\partial \phi ^{2}}+\frac{\partial
^{2}V_{SYM}(\phi )}{\partial \phi \text{ }\partial u}\right] _{\phi =u=v} 
\nonumber \\
&=&-C_{\xi }(v)\left[ \frac{\partial ^{2}V_{SYM}(\phi )}{\partial \phi ^{2}}%
\right] _{\phi =u=v}+\xi \frac{\partial v}{\partial \xi }(1-C_{u}(v)/v)%
\left[ \frac{\partial ^{2}V_{SYM}(\phi )}{\partial \phi ^{2}}\right] _{\phi
=u=v}=0,
\end{eqnarray}
from which we obtain 
\begin{equation}
\xi \frac{\partial v}{\partial \xi }=\frac{C_{\xi }(v)}{1-C_{u}(v)/v}.
\end{equation}
so that the gauge dependence of the VEV is different from that of $%
\overline{R_{\xi }}$ gauge given in Eq.(13). Since the $C_{u}(v)$
dependent terms emerges from the one loop, the result of Eq.(36) will be
different from the property(ii) beyond the two loop order. The total derivative of the
effective potential $V(\phi )$ at $\phi =v$ with respect to $\xi $ becomes 
\begin{equation}
\xi \frac{d}{d\xi }V_{BS}(v)=\xi \frac{\partial v}{\partial \xi }\left[ 
\frac{\partial V_{SYM}}{\partial \phi }+\frac{\partial V_{SYM}}{\partial u}%
\right] _{\phi =u=v}-C_{\xi }(v)\left[ \frac{\partial V_{SYM}}{\partial \phi 
}\right] _{\phi =u=v}=0.
\end{equation}
By using $x=u$ case of Eq.(18) and Eq.(20), we can see that the effective
potential is gauge independent when $\phi =v $ as in case of $\overline
{R_{\xi }}$ gauge. 

The gauge dependence of the VEV played an important role in showing the 
gauge independence of the pole mass in case of the $\overline{R_{\xi }}$
and since it has changed as in Eq.(36) in case of the $R_{\xi}$ gauge, it will be necessarily
to investigate whether the pole mass remains gauge independent in this case.
It was shown that the one-loop pole mass is gauge
invariant in $R_{\xi }$ gauge Ref.[9,10,15]
and let us consider the problem of the gauge independence of the higher
order pole mass in the broken symmetry phase of the $R_{\xi }$ gauge. By
applying $\xi \frac{\partial }{\partial \xi }$ to the defining equation of
the pole mass in the broken symmetry phase as in Eq.(16) and by using
Eqs.(6),(15) and (36), we obtain 
\begin{eqnarray}
0 &=&\xi \frac{\partial }{\partial \xi }\left[ \frac{\partial ^{2}\Gamma
_{SYM}}{\partial \phi ^{2}}\right] _{\phi =u=v,\text{ }p^{2}=-M^{2}}=\xi 
\frac{\partial v}{\partial \xi }\left[ \frac{\partial ^{3}\Gamma _{SYM}}{%
\partial \phi ^{3}}+\frac{\partial ^{3}\Gamma _{SYM}}{\partial \phi
^{2}\partial u}\right] _{\phi =u=v,\text{ }p^{2}=-M^{2}}+\xi \frac{\partial
M^{2}}{\partial \xi }\left[ \frac{\partial ^{3}\Gamma _{SYM}}{\partial \phi
^{2}\partial (p^{2})}\right] _{\phi =u=v,\text{ }p^{2}=-M^{2}}  \nonumber \\
&&-\left[ \frac{\partial ^{2}C_{\xi }(\phi )}{\partial \phi ^{2}}\text{ }%
\frac{\partial \Gamma _{SYM}}{\partial \phi }+2\text{ }\frac{\partial C_{\xi
}(\phi )}{\partial \phi }\text{ }\frac{\partial ^{2}\Gamma _{SYM}}{\partial
\phi ^{2}}+C_{\xi }(\phi )\text{ }\frac{\partial ^{3}\Gamma _{SYM}}{\partial
\phi ^{3}}\right] _{\phi =u=v,\text{ }p^{2}=-M^{2}}  \nonumber \\
&=&\xi \frac{\partial v}{\partial \xi }\left[ \frac{\partial ^{3}\Gamma
_{SYM}}{\partial \phi ^{3}}-\frac{1}{u}(\frac{\partial ^{2}C_{u}(\phi )}{%
\partial \phi ^{2}}\text{ }\frac{\partial \Gamma _{SYM}}{\partial \phi }+2%
\text{ }\frac{\partial C_{u}(\phi )}{\partial \phi }\text{ }\frac{\partial
^{2}\Gamma _{SYM}}{\partial \phi ^{2}}+C_{u}(\phi )\text{ }\frac{\partial
^{3}\Gamma _{SYM}}{\partial \phi ^{3}})\right] +\xi \frac{\partial M^{2}}{%
\partial \xi }\left[ \frac{\partial ^{3}\Gamma _{SYM}}{\partial \phi
^{2}\partial (p^{2})}\right] _{\phi =u=v,\text{ }p^{2}=-M^{2}}  \nonumber \\
&&-\left[ \frac{\partial ^{2}C_{\xi }(\phi )}{\partial \phi ^{2}}\text{ }%
\frac{\partial \Gamma _{SYM}}{\partial \phi }+2\text{ }\frac{\partial C_{\xi
}(\phi )}{\partial \phi }\text{ }\frac{\partial ^{2}\Gamma _{SYM}}{\partial
\phi ^{2}}+C_{\xi }(\phi )\text{ }\frac{\partial ^{3}\Gamma _{SYM}}{\partial
\phi ^{3}}\right] _{\phi =u=v,\text{ }p^{2}=-M^{2}}  \nonumber \\
&=&\xi \frac{\partial v}{\partial \xi }(1-\frac{C_{u}(v)}{v})\left[ \frac{%
\partial ^{3}\Gamma _{SYM}}{\partial \phi ^{3}}\right] _{\phi =u=v,\text{ }%
p^{2}=-M^{2}}+\xi \frac{\partial M^{2}}{\partial \xi }\left[ \frac{\partial
^{3}\Gamma _{SYM}}{\partial \phi ^{2}\partial (p^{2})}\right] _{\phi =u=v,%
\text{ }p^{2}=-M^{2}}-C_{\xi }(v)\left[ \text{ }\frac{\partial ^{3}\Gamma
_{SYM}}{\partial \phi ^{3}}\right] _{\phi =u=v,\text{ }p^{2}=-M^{2}}.
\end{eqnarray}
By using Eq.(36), we can see that $\xi \frac{\partial M^{2}}{\partial \xi }$
vanishes and hence pole mass is a gauge independent in broken symmetry phase
of the $R_{\xi }$ gauge.

In conclusion, we have investigated the difference in choice of the gauge
whether the Lagrangian density have parameter that has any relation to the
VEV in the $\overline{MS}$ scheme. In case of the $\overline{R_{\xi }}$ gauge where the Lagrangian
density do not have parameter which has a relation to the VEV, the gamma
function of VEV is same as that of the Higgs field and the pole mass is
gauge independent. In case of the broken symmetry phase of the $R_{\xi }$
gauge, the gamma function of the VEV is different from that of the Higgs
field due to the presence of the parameter in
the gauge function that should be identified as a VEV in the broken symmetry
phase. Although this was known previously\cite{VEV-1}\cite{VEV-2}, our
result is different from their results beyond the two loop order.
The gauge parameter dependence of the VEV also needs modification from the property(ii)
. However the pole mass remains gauge independent.


\begin{thebibliography}{99}
\bibitem{R-ksi}  K. Fujikawa, B.W. Lee and A.I. Sanda, Phys. Rev. D6 (1972)
2923; Y.P. Yao, Phys. Rev. D7 (1973) 1647.

\bibitem{Nielsen}  N. K. Nielsen, Nucl. Phys. B101 (1975) 173.

\bibitem{VEV-1}  M. Sperling, D. Stockinger, A.Voigt, JHEP 1307 (2013) 132.

\bibitem{VEV-2}  M. Sperling, D. Stockinger, A.Voigt, JHEP 1401 (2014) 068.

\bibitem{Kas}  B. Kastening, Phys. Rev. D51\textbf{\ }, (1995) 265.

\bibitem{Kim1}  C. Kim, Phys. Rev. D90,067701 (2014).

\bibitem{RG}  C. Ford, D. R. T. Jones, P. W. Stephenson and M. B. Einhorn,
Nucl. Phys. B\textbf{395} (1993) 17.

\bibitem{Kim2}  C. Kim, J. Kor. Phys. Soc. 62,1097 (2013).


\bibitem{Fraser}  I. J. R. Aitchison and C. M. Fraser, Ann. Phys. 156,
(1984) 1.

\bibitem{Zeit}  A. F. de lima and D. Bazeia, Zeitschrift fur Phys. C45,
(1990) 471.

\bibitem{Willey}  W. Loinaz and R. S. Willey, Phys. Rev. D56, (1997) 7416.

\bibitem{integral}  See for example, M. E. Peskin and D. V. Schroeder, , 
\emph{An Introduction to Quantum Field Theory} (Addison-Wesley, Reading,
1995).

\bibitem{RG-1}  M. E. Machacek and M. T. Vaughn, Nucl. Phys. B \textbf{222}
(1983) 83.

\bibitem{RG-2}  M. E. Machacek and M. T. Vaughn, Nucl. Phys. B \textbf{249}
(1985) 70.

\bibitem{Johnson}  D. Johnston, Nucl. Phys. B \textbf{253} (1985) 687.
\end{thebibliography}
\end{document}